\documentclass{iaus}
\usepackage{graphicx}
\usepackage{natbib}
\def\mbh{{$\mathcal M_{\rm BH}$}}

\def\mbb{{$\mathcal M_{\rm BH}-\mathcal M_{\rm bulge}$}}
\def\msig{{$\mathcal M_{\rm BH}-\sigma_*$}}
\def\mbt{{$\mathcal M_{\rm BH}-\mathcal M_{\rm total}$}}

\def\mbulge{{$\mathcal M_{\rm bulge}$}}

\def\mgal{{$\mathcal M_{\rm gal}$}}
\def\msol{{$\mathcal M_\odot$}}

\def\mbhmerge{{$\mu_{\rm merge}$}}
\def\phigal{{$\Phi({\mbox \mbulge})$}}
\def\phimbh{{$\Phi({\mbox \mbh})$}}
\def\lqso{{$L_{\rm QSO}$}}
\def\condbhbul{{P(\mbh$|$\mbulge)}}
\def\condbulbh{{P(\mbulge$|$\mbh)}}
\def\pcorr{{\rm{P(}\mbh$,$\mbulge\rm{)}}}
\def\avgmbh{{$\left<\mbox{\mbh}\right>$}}
\def\avgmbul{{$\left<\mbox{\mbulge}\right>$}}

\title[Statistics of \mbb\ Correlation and Co-evolution] {What Do Statistics
Reveal About the \mbb\ Correlation and Co-evolution?}

\author[Chien Y. Peng] {Chien Y. Peng$^1$}

\affiliation{
  $^1$ Herzberg Institute of Astrophysics, National Research Council of Canada\\
  5071 West Saanich Road, Victoria, British Columbia, V9E 2E7, Canada\\ 
  email: {\tt cyp@nrc-cnrc.gc.ca}
}

\pubyear{2009}
\volume{267}  %% insert here IAU Symposium No.
\pagerange{1-8}
\date{January 27, 2010}
\jname{Co-Evolution of Central Black Holes and Galaxies}
\editors{B.M.\ Peterson, R.S.\ Somerville, \& T.\ Storchi-Bergmann, eds.}
\begin{document}

\maketitle

\begin{abstract} Observational data show that the correlation between
supermassive black holes (\mbh) and galaxy bulge (\mbulge) masses follows a
nearly linear trend, and that the correlation is strongest with the bulge
rather than the total stellar mass (\mgal).  With increasing redshift, the
ratio $\Gamma=$ \mbh/\mbulge\ relative to $z=0$ also seems to be larger for
\mbh\ $\gtrsim10^{8.5}$ \msol.  This study looks more closely at statistics to
better understand the creation and observations of the \mbb\ correlation.  It
is possible to show that if galaxy merging statistics can drive the
correlation, minor mergers are responsible for causing a convergence to
linearity most evident {\it at high masses}, whereas major mergers have a
central limit convergence that more strongly {\it reduces the scatter}.  This
statistical reasoning is agnostic about galaxy morphology.  Therefore,
combining statistical prediction (more major mergers $\Longrightarrow$ tighter
correlation) with observations (bulges = tightest correlation), would lead one
to conclude that more major mergers ({\it throughout an entire merger tree},
not just the primary branch) give rise to more prominent bulges.  Lastly, with
regard to controversial findings that $\Gamma$ increases with redshift, this
study shows why the luminosity function (LF) bias argument, taken correctly at
face value, actually strengthens, rather than weakens, the findings.  However,
correcting for LF bias is unwarranted because the BH mass scale for quasars is
bootstrapped to the \msig\ correlation in normal galaxies at $z=0$, and
quasar-quasar comparisons are mostly internally consistent.  In Monte-Carlo
simulations, high $\Gamma$ galaxies are indeed present: they are statistical
outliers (i.e.  ``under-merged'') that take longer to converge to linearity
via minor mergers.  Another evidence that the galaxies are undermassive at
$z\gtrsim2$ for their \mbh\ is that the quasar hosts are very compact for
their expected mass.


\end{abstract}

\firstsection % if your document starts with a section,
\section{Introduction} 

The discoveries of fundamental correlations between \mbh\ with stellar
velocity dispersion \citep{ferrarese00, gebhardt00a} and \mbulge\
\citep{kormendy95, magorrian98} have led to modern views that supermassive
black hole activities may have a strong impact on galaxy evolution
\citep[e.g][]{kauffmann00, granato04, dimatteo05}.  The correlation between
\mbh\ and \mbulge\ is remarkable in that it is almost linear, has only a
scatter of 0.3 dex, and holds true over 5 orders of magnitude in \mbh\ dynamic
range.  Locally, the ratio of \mbulge/\mbh\ $\approx 800$ \citep{marconi03,
haering04}.

How did the correlations come about and how do selection biases affect our
observations of the correlations?  Direct cause and effect are not only
possible, there are numerous theoretical proposals.  While quasar feedback is
one of the most investigated and favored explanations \citep{dimatteo05,
robertson06a, hopkins07a}, galaxy mergers may perhaps share a role.  This
study therefore isolates the role of merger statistics to examine how it might
affect the growth of the \mbb\ correlation.  It also examines more closely how
luminosity selection biases affect the inference of \mbb\ correlation and its
evolution since $z\approx 4$, as inferred using quasars.

%%%%%%%%%%%%%%%%%%%%%%%%%%%%%%%%%%%%%%%%%%%%%%%%%%%%%%%%%%%%%%%%%%%%%%%%%%%%%

\begin{figure}[t]
    % \vspace*{-2.0 cm}
    \begin{center}
        \includegraphics[width=5in]{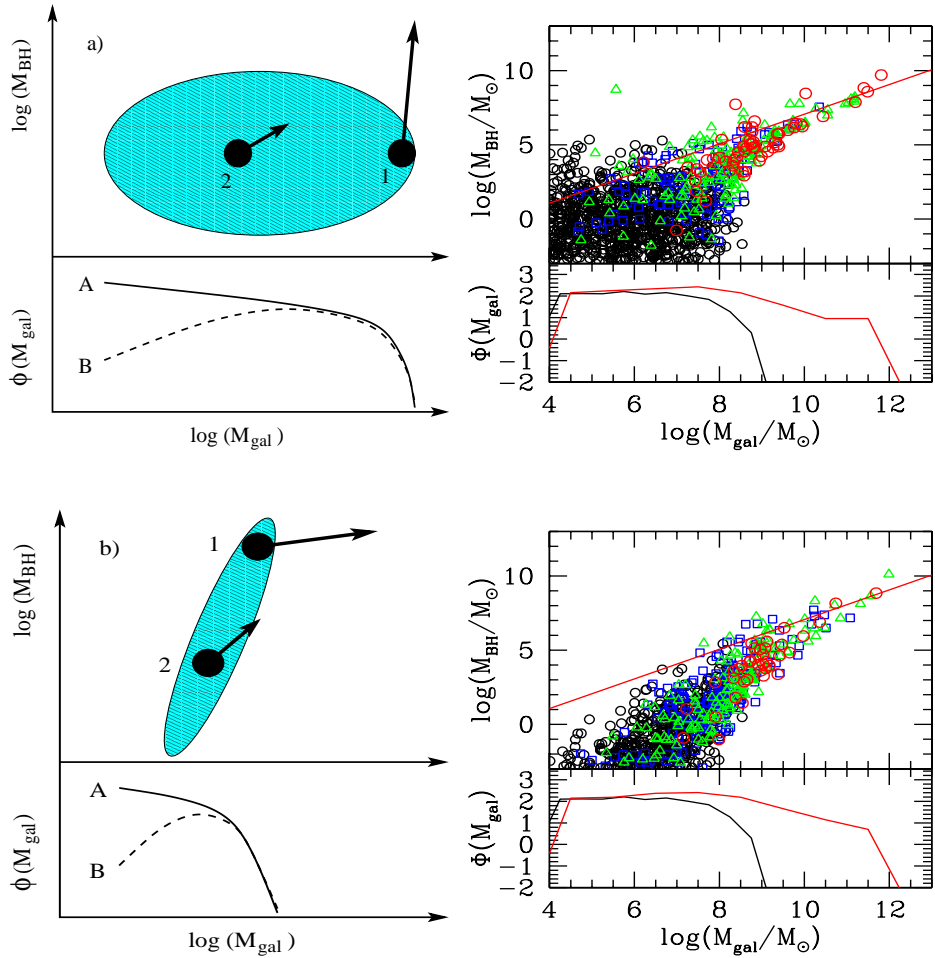} 
        % \vspace*{-1.0 cm}
    \end{center}

    \caption{{\it How mergers cause \mbh\ and \mgal\ to correlate.} {\it a})
	No correlation initially.  Objects at location 1 increases \mgal\ more
	quickly than \mbh\ upon merging, compared to objects at location 2.
	The correlation steepens over time due to minor mergers.  {\it b})
	Steep correlation initially.  The opposite situation occurs, i.e.
	\mgal\ grows more quickly than \mbh\ for objects at 1 compared to 2.
	Symmetry between scenarios {\it a} and {\it b} produces linear,
	relation, asymptotically.  Panels on the {\it right} are Monte-Carlo
	simulations with the initial conditions shown on {\it left}.}

    \label{fig:sims}
\end{figure}

\begin{figure}[ht]
    \begin{center}
        \includegraphics[width=5in]{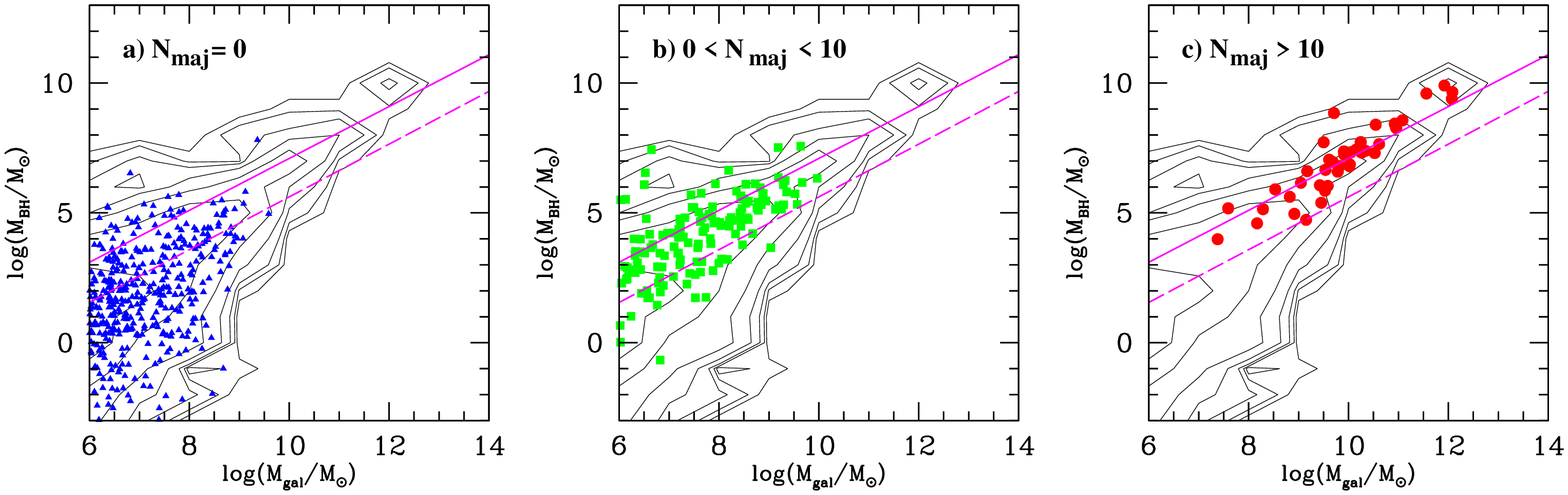} 
        % \vspace*{-1.0 cm}
    \end{center}

    \caption{{\it Effects of major mergers based on Monte-Carlo simulations.}
	The solid line corresponds to a linear correlation; it is not a fit.
	The contours show the distribution of points at the end of the Monte
	Carlo simulation.  {\it a}) No major mergers $N_{\rm maj} = 0$.  The
	dashed line shows roughly the mean of the points with slope fixed to a
	linear correlation.  {\it b}) Cumulative major mergers $0 < N_{\rm
	maj} < 10$ over all progenitors in an merger tree of a galaxy.  {\it
	c}) Cumulative major mergers $N_{\rm maj} \ge 10$ over all progenitors
	galaxies.}

    \label{fig:major}

    \vspace{-0.05truein}

\end{figure}

\section {Merging Statistics: How \mbb\ Correlation Can Result and Why
it is Important to Also Consider the \mbt\ Relation}

How galaxy merging affects the BH vs. bulge correlations has been considered
by several studies \citep[e.g.][]{islam03, ciotti01} from a purely statistical
standpoint, and using specific initial conditions and assumptions (e.g.  no
scatter or starting with a prior correlation).  Going further,
\citet{peng07} shows the two most salient features of the
\mbb\ correlation:  linearity and strong correlation with bulges, can be
attained without having to make assumptions about the initial conditions.  The
heuristic toy model proposed by \citet{peng07}, shown in Fig.~\ref{fig:sims},
explains that a linear quality of the \mbb\ relation comes about because of
minor mergers.  Minor mergers affect the \mbh\ -- \mgal\ relation in a way that
over some number of events would attain a cosmic average ratio in \mbh/\mgal.
That there is necessarily a correlation can be reasoned from symmetry
argument; that the correlation trends toward {\it linearity} can be understood
by noticing that the only way minor mergers can no longer change the
\mbh/\mgal\ ratio is when it is the same value everywhere along the mass
sequence.

Major mergers, however, play a different role: they do not significantly
change the ratio of \mbh/\mgal\ because the galaxies involved in merging have
both similar $\left<\mbox{\mbh}\right>$ and $\left<\mbox{\mgal}\right>$ by
definition of major mergers.  As explained in more detail in \citet{peng07},
in the limit where major mergers are occurring between identical mass
galaxies, the scatter of log(\mbh) after merging roughly goes like $
     \sigma\left(\mbox{log}\left(\mbox{\mbhmerge}\right)\right) =
        \frac{\sigma\left<\mbox{\mbh$_{,1}$} + 
        \mbox{\mbh$_{,2}$}\right>}{\left<\mbox{\mbh$_{,1}$} + \mbox{\mbh$_{,2}$}\right>},
$ due to the central limit theorem.  Thus, because the sum,
$\left<\mbox{\mbh$_{,1}$} + \mbox{\mbh$_{,2}$}\right>$, increases more quickly
than the dispersion, $\sigma\left<\mbox{\mbh$_{,1}$} +
\mbox{\mbh$_{,2}$}\right>$, the scatter in the \mbh\ distribution after
merging, $\sigma\left(\mbox{log}\left(\mbox{\mbhmerge}\right)\right)$,
decreases with each major merger.  The effect of central limit convergence due
to major mergers can be quite dramatic.  Fig.~\ref{fig:major} shows results
from one possible Monte-Carlo simulation, in which there is no correlation
initially, and has 2 orders of magnitude in scatter.  Minor mergers
(Fig.~\ref{fig:major}{\it a}) alone do not reduce the scatter significantly by
the end of the simulation.  However, major mergers cause a rapid decrease in
scatter in only a few events.  Note that the relevant accounting of major
mergers is the cumulative sum {\it over the entire merger tree}, i.e. over all
progenitors, their progenitors, etc., rather than the more common approach of
tracking the main branch.

It is worth noting that statistical reasoning does not predict morphology from
first principles.  Therefore statistical reasoning can explain the
observations of a tight \mbb\ correlation if and only if massive bulges were
preferentially formed through more major mergers than disky galaxies, summed
over all progenitor histories.  Even though the notion that major mergers lead
to formation of bulges is now widely regarded to be true, it is interesting
that it can be reasoned purely from statistical principles and the known
existence of a tight \mbb\ correlation.  Furthermore, the \mbb\ relation may
be a special case of the \mbt\ relation, despite the latter having a larger
scatter.  Thus, to understand the co-evolution of galaxies with \mbh, one
ought to consider both the \mbb\ and \mbt\ relations.

%%%%%%%%%%%%%%%%%%%%%%%%%%%%%%%%%%%%%%%%%%%%%%%%%%%%%%%%%%%%%%%%%%%%%%%%%%%%%

\begin{figure}[ht]
    % \vspace*{-2.0 cm}
 
    \begin{center}
        \includegraphics[width=5in]{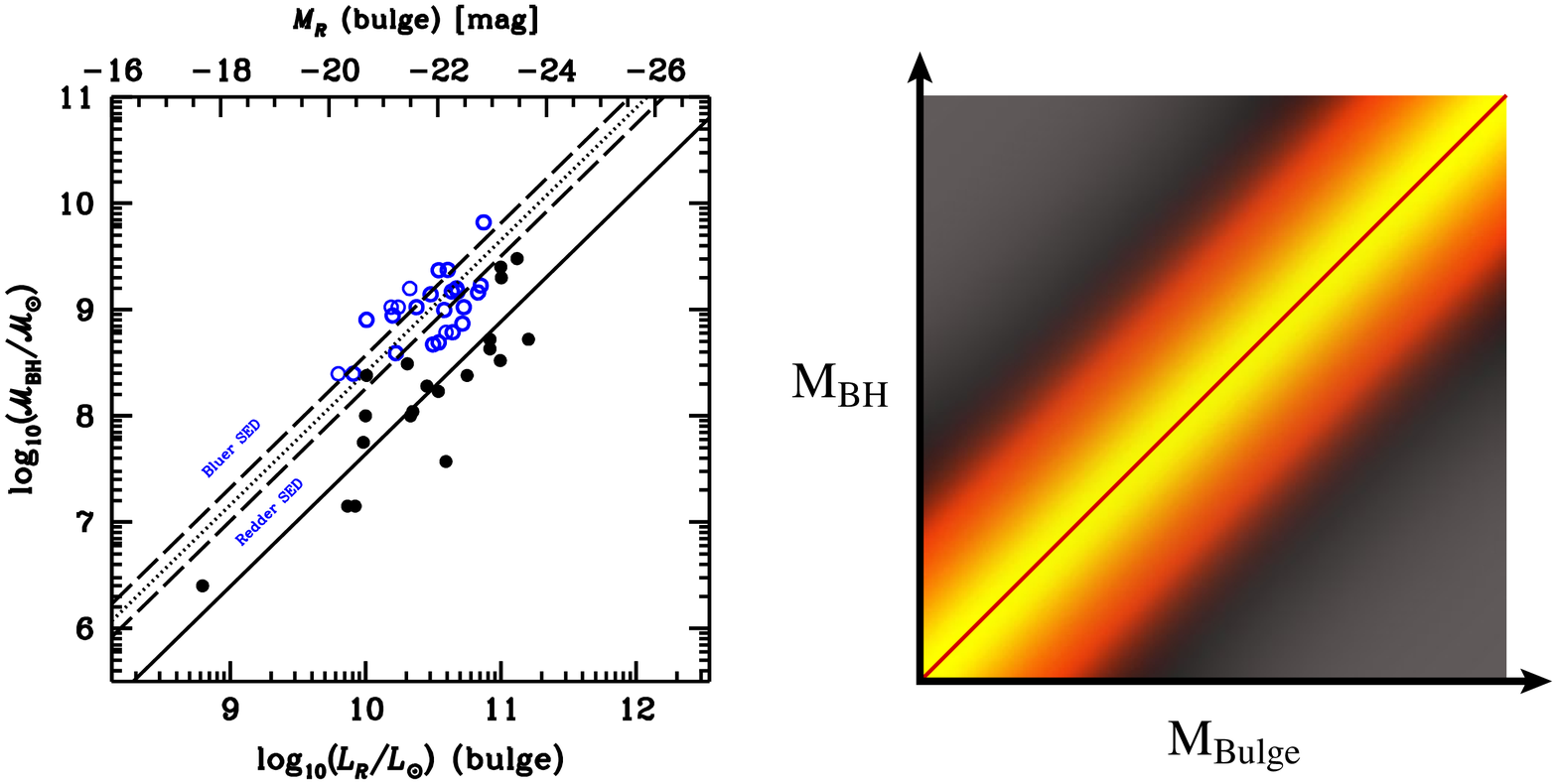} 
        % \vspace*{-1.0 cm}
    \end{center}

    \caption{Observed $z\gtrsim 1.7$ quasar data (open circles) from
	\citet{peng06b}.  Solid circles are $z=0$ normal galaxies.  The
	absolute luminosities are how the galaxies would appear after
	accounting for luminosity fading.}

    \label{fig:highz}

    \caption{{\it The intrinsic correlation.} What it means for the intrinsic
	correlation between \mbh\ and \mbulge\ to be linear.  This
	distribution is also called ``the prior'' and the conditional
	\condbhbul. To simplify discussion, we assume \condbhbul=\condbulbh\
	as shown; doing so does not affect the conclusion qualitatively.}

    \label{fig:intrinsic}

\end{figure}

\section {Luminosity Function Bias of Galaxies, Quasars and Other Biases}

In recent years, there have been a number of efforts to study the \mbb\
correlation beyond the local universe using quasars, radio galaxies, and other
means \citep[e.g.][]{mclure06, peng06b, peng06a, woo06, treu07}.  For the most
part, the studies find that the central BHs were larger at $z\gtrsim 2$ in the
past for a given bulge stellar mass by a factor $\Gamma$ of $3
\lesssim \Gamma \lesssim6$ \citep{peng06b, peng06a}, shown in
Fig.~\ref{fig:highz}.  These findings have been called into question by other
studies on the basis that the luminosity function (LF) selection was not
explicitly accounted \citep[e.g.][]{lauer07b}.  However, the criticisms have
not been very germane, both because the BH mass scale in quasars is not on an
absolute scale, and the LF bias goes in the opposite direction in quasars than
claimed, as discussed below.  The issues are subtle and have led to
substantial confusion.

\begin{figure}

%   \vspace*{1.0 cm}

    \begin{center}
        \includegraphics[width=5.3in]{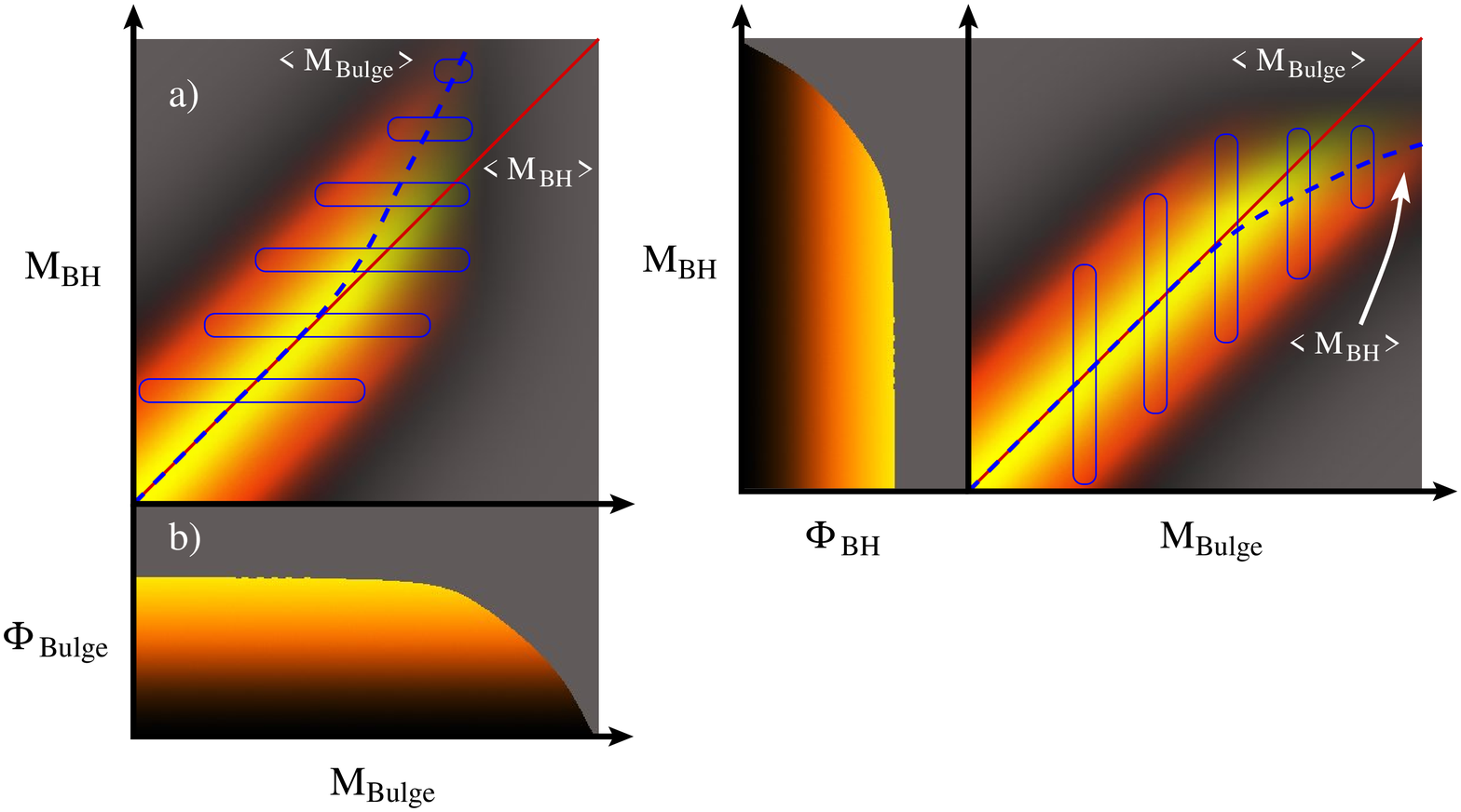} 
        % \vspace*{-1.0 cm}
    \end{center}

    \caption{{\it Measuring BHs in normal galaxies.} {\it a}) Given
	the intrinsic prior of Fig.~\ref{fig:intrinsic}, what is actually
	observed when objects are first drawn from the bulge mass (i.e.
	luminosity) function $\Phi_{\rm Bulge}$ (Panel~{\it b}), followed by
	perfect \mbh\ measurement.  Note that $\left< {\mbox \mbulge} \right>$
	(with \mbh\ as the independent variable, as represented by horizontal
	rectangles as visual aid) trends upward (dashed line) even though
	$\left< {\mbox \mbh} \right>$ (solid line, with \mbulge\ as the
	independent variable) is not biased.}

    \label{fig:galsel}

    \caption{{\it Measuring galaxies around luminous quasars.} Selecting
	quasars from the quasar mass (i.e. luminosity) function $\Phi_{\rm
	BH}$, given that the intrinsic correlation P(\mbulge$|$\mbh) between
	\mbh\ and \mbulge\ is linear (Fig.~\ref{fig:intrinsic}).  The tapering
	of the correlation at high \mbh\ is due to there being fewer luminous
	quasars in the universe, as illustrated by the mass function \phimbh\
	to the left.  Note that $\left< {\mbox \mbh} \right>$ (at a given
	\mbulge) trends to the right (dashed line), but $\left< {\mbox
	\mbulge} \right>$ (solid line) is not biased.}

    \label{fig:qsosel}
\end{figure}

\subsection {Revisiting the Luminosity Function Selection Bias to See Why It
Affects the \mbb\ Correlation in Galaxies Differently From Quasars.}

Figures ~\ref{fig:intrinsic}--\ref{fig:qsosel} illustrate schematically the
idea of the LF selection bias.  Figure~\ref{fig:intrinsic} is the prior that
there is an intrinsic, perfectly linear, correlation between \mbh\ and
\mbulge.  This intrinsic correlation \condbhbul\ is also known as the
conditional.  Even though a linear relation does not require \condbhbul\ to be
the same as \condbulbh, doing so in the discussion below does not affect the
directional sense of the conclusions.

The {\it luminosity function bias} was pointed out at least as early as
\citet{adelberger05}, and more recently by \citet{fine06, salviander06,
lauer07b}.  In essence, the act of selecting a sample of galaxies to observe
leaves an imprint of the LF on the correlation of \mbh\ vs. \mbulge.  To
obtain the \mbb\ correlation in normal galaxies, the observing sequence is to
first select bulges or galaxies from \phigal, the galaxy bulge mass function
shown schematically in Fig.~\ref{fig:galsel}{\it b}, followed by measuring the
BH through stellar dynamics or other means.  The latter probability:
measuring a BH of mass \mbh\ {\it after selecting on
\mbulge}, is the conditional \condbhbul\ of Fig.~\ref{fig:intrinsic}.  The
observational sequence: \phigal\ $\times$ \condbhbul\ therefore establishes
the observed correlation \pcorr\ shown in Fig.~\ref{fig:galsel}{\it a}.  The
effect of selecting on \phigal\ tapers off the underlying correlation at the
right side indiscriminantly of \mbh.  This LF imprint is present even if every
galaxy and BH can be detected and measured precisely.  It is {\it not} a
Malmquist bias, and the effect prevents us from directly observing the
intrinsic correlation.  Fig.~\ref{fig:galsel}{\it a} is the same, in essence,
as Fig.~2 in \citet{lauer07b}.  The tapering by \phigal\ causes the \avgmbul\
for a given \mbh\ to deviate from the intrinsic trend, as illustrated by the
dashed line in Fig.~\ref{fig:galsel}{\it a}.  It leads to the notion that ``at
a given BH mass, there are more low mass host galaxies than high mass.'' This
is the main reason behind the argument that high-$z$ data in
Fig.~\ref{fig:highz} are biased.

However, what is subtle and widely misconstrued is the fact that the
distribution of Fig.~\ref{fig:galsel}{\it a} applies only to normal galaxies
but not for quasars.  In quasars, the reverse observational sequence occurs,
i.e.  {\it measuring host galaxies around BHs}, as opposed to {\it measuring
BHs in normal galaxies}.  When selecting on quasars, there is an agreement
that one does not draw them from the galaxy luminosity function, but instead
from $\Phi(L_{\rm QSO})$ -- the quasar luminosity function.  Moreover, because
\mbh\ in quasars scales like \lqso$^{0.5}$ \citep{kaspi00}, and quasars appear
to radiate at fixed fraction of Eddington ratio \citep{kollmeier06}, selecting
on \lqso\ is essentially drawing on \phimbh (Fig.~\ref{fig:qsosel}, left).
After selecting on quasars, the host galaxy masses \mbulge\ are then drawn
from the conditional probability of finding \mbulge\ around a BH of mass \mbh,
i.e. \condbulbh.  Therefore, the observational sequence for quasars is given
by the product: \pcorr = \phimbh\ $\times$ \condbulbh.  Comparing this to
\pcorr\ for normal galaxies, one notices the labels of \mbh\ and \mbulge\ are
simply switched.  Doing so leads to Fig.~\ref{fig:qsosel}, whereby the
\phimbh\ selection attenuates the intrinsic correlation of
Fig.~\ref{fig:intrinsic} on the upper (\mbh) side.  In other words, in quasars
it is \avgmbh, not \avgmbul, which is lower than intrinsic.  Therefore, the
fact that high-$z$ data in Fig.~\ref{fig:highz} lie on the opposite side of
the expected trend is a testament to a positive evolution in $\Gamma$ if the
BH mass scale is absolute.  However, it is not, as discussed below, which
means this effect is only secondary.

\begin{figure}[t]
    % \vspace*{-2.0 cm}
    \begin{center}
        \includegraphics[width=5.3in]{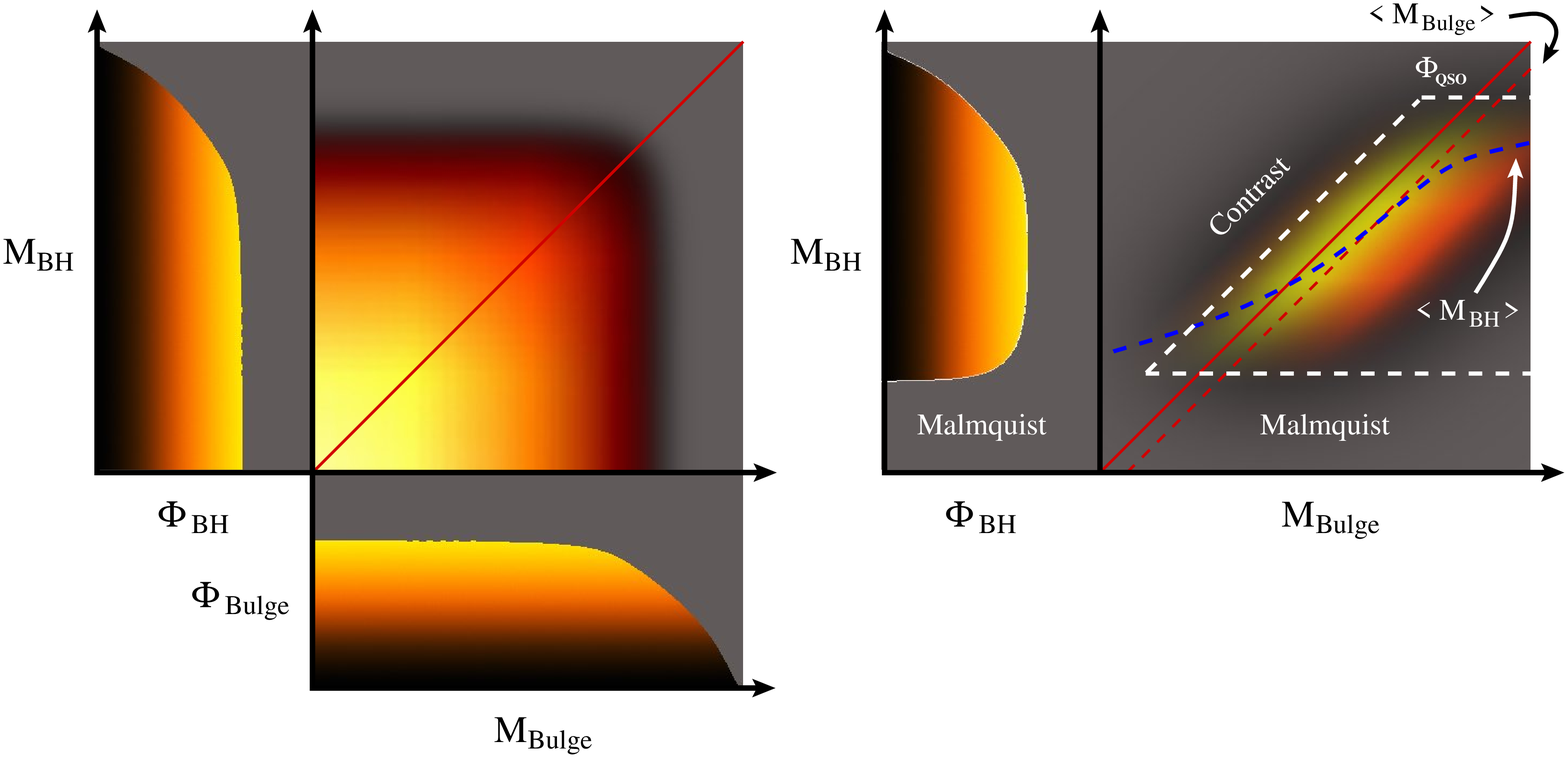} 
        % \vspace*{-1.0 cm}
    \end{center}

    \caption{{\it The joint distribution \pcorr$ = $\phigal $\times$\phimbh.}
	A common, but wrong, notion that, for every quasar observed from the
	BH mass function ($\Phi_{\rm BH}$), the host galaxies can thereafter
	be drawn from the bulge mass function $\Phi_{\rm Bulge}$ = \condbulbh.
	Rather than producing the \mbb\ correlation, the joint product results
	in no correlation.}

    \label{fig:wrong}

    \caption{{\it Other selection biases affecting quasar observations.}
	Observing the \mbb\ correlation in quasars has additional biases
	shown, given that the intrinsic correlation P(\mbulge$|$\mbh) between
	\mbh\ and \mbulge\ is linear (Fig.~\ref{fig:intrinsic}).  The tapering
	of the correlation at high \mbh\ is again due to selection on \phimbh.
	Malmquist bias selects against faint quasars, independent of \mbulge\
	(lower dashed line).  On the other hand, faint host galaxies sitting
	beneath luminous quasars are hard to detect, giving rise to a diagonal
	selection bias.  The exact angle of the diagonal bias depends on the
	degree of host galaxy magnification.  Both \avgmbh\ and \avgmbul\ are
	now shifted to the right of nominal center (solid line) due to
	contrast bias.  Comparing this expected distribution with
	Fig.~\ref{fig:highz} shows a lack of LF bias in high-$z$
	observations.}

    \label{fig:qsoall}
\end{figure}

It appears one reason there is widespread misconception on this issue is a
tendency to apply the intuitive notion that there are more low mass than high
mass galaxies, when doing so is not appropriate.  In other words, after
selecting a quasar from the BH mass function \phimbh, the tendency is to
believe the host galaxy should be drawn from the galaxy mass function,
\condbulbh = \phigal, because there are more low mass than high mass galaxies,
as opposed to the intrinsic correlation of Fig.~\ref{fig:intrinsic}.  Doing so
leads to the joint product \pcorr$ = $\phigal $\times$\phimbh\ which
heuristically produces a distribution given by Fig.~\ref{fig:wrong}.  Clearly,
observations do not support this because the joint product produces no
correlation between \mbh\ and \mbulge.  Note that \citet{lauer07b} did not
make this particular error; their conditional probability comes from the
linear correlation of Fig.~\ref{fig:intrinsic}, not \phimbh.  Statistically,
the only way for the conditional \condbulbh = \phigal\ is if \mbh\ and
\mbulge\ are intrinsically unrelated.

\subsection {The Effects of Malmquist and Quasar-to-Host Galaxy Contrast
Biases on the \mbb\ Correlation in Quasars.}

Malmquist bias is another common factor used to argue against findings that
the ratio of \mbh/\mbulge\ is higher at high-$z$ than now.  However,
Fig.~\ref{fig:qsoall} illustrates schematically that Malmquist bias only
attenuates the underside of the distribution.  It does not affect the trend at
the massive end.  It is also qualitatively very different from high-$z$
observations of Fig.~\ref{fig:highz} because, as seen in
Fig.~\ref{fig:qsoall}, the attenuation is uniform at a constant \mbh; it does
not cause the points to lie systematically to the left of the correlation
line.

Lastly, measuring host galaxies around quasars is affected by the fact that
only luminous host galaxies can be detected from beneath luminous quasars.
Under normal circumstances without gravitational lensing, host galaxies of
quasars are extremely difficult to detect when the quasar:host ratio is larger
than $10:1$ at a seeing of $0.1$ arcsec.  This selection bias tapers the
correlation along a diagonal line illustrated schematically in
Fig.~\ref{fig:qsoall}; the angle of the diagonal depends on the magnification
ratio if the quasar sample is from gravitational lenses, as in
Fig.~\ref{fig:highz}.  Nevertheless, the observational pressure is to shift
the \avgmbh\ and \avgmbul\ averages to the right of the intrinsic correlation.

Comparing Fig.~\ref{fig:qsoall} with Fig.~\ref{fig:highz} therefore
qualitatively illustrates that the finding of a larger \mbh/\mbulge\ ratio in
high-$z$ quasars is not due to known luminosity selection effects.
Qualitatively, observational pressures greatly favor galaxy detections to the
right of the correlation line where the quasar luminosity contrast is low.
The missing objects to the right of the intrinsic correlation may be caused by
quasar surveys that fail to classify low contrast, thus redder, objects as
being quasar candidates.  However, given that even redder and lower contrast
systems make it into the \citet{peng06b} quasar sample at $z=1$, this effect
is judged on face value to probably not be the main culprit.

Note that, hypothetically, it is possible for studies using {\it other
selection functions} beside ones mentioned to distill a sample of low
luminosity quasars that are then found to the right of the correlation.  That
would not necessarily contradict current conclusions using quasars.  Instead,
that hypothetical sample can have properties that distinguish them physically
from the host galaxies of luminous quasars.  Selection functions that draw on
different physical attributes may find objects in different parameter space of
the same underlying \mbt\ correlation.  This might explain the different
conclusions seen between quasars and sub-mm galaxies hosting active nuclei
\citep[e.g][]{alexander08}.  To talk about evolution, it is therefore
necessary to compare objects selected based on the {\it same physical and
observational} selection functions.  In that respect, the quasar-quasar
comparisons of high-$z$ and low-$z$ are currently the most {\it internally
consistent} sample to address the issue of the \mbb\ evolution.  Lastly it is
important to note that where selection biases strongly partition observable
parameter spaces, it is important to not only consider the mean of some trend,
but also the distribution {\it as a whole.} Taking out biases in distributions
from known selection functions are feasible.

\begin{figure}[t]
    \begin{center}
        \includegraphics[width=3.9in]{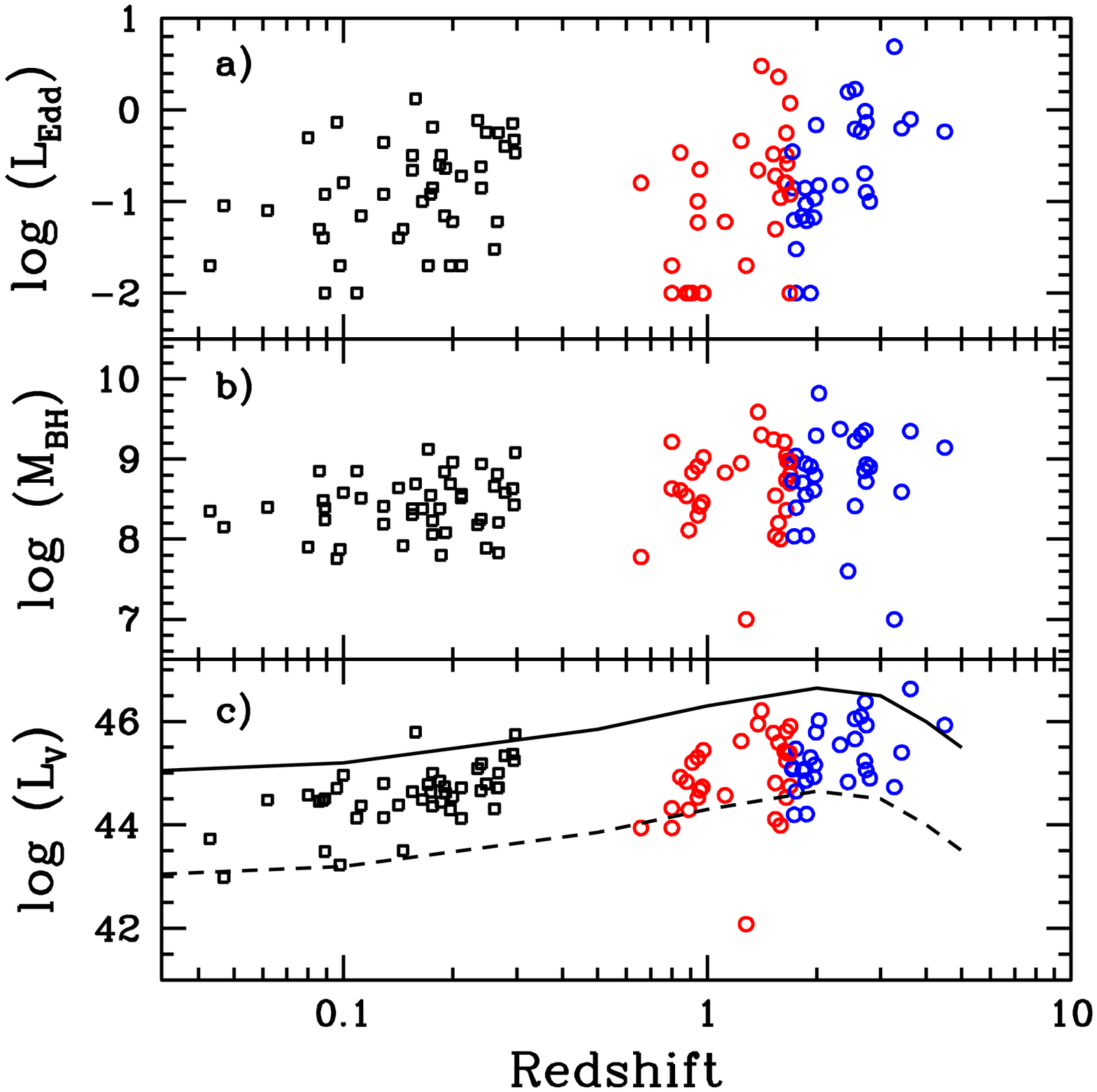} 
    \end{center}

    \caption{{\it Comparison of quasar properties at high and low $z$.} {\it
	a}) Quasar radiating efficiency in units of [$L_{\rm QSO} / L_{\rm
	Eddington}]$.  {\it b}) \mbh\ in the quasar sample in [\msol].  {\it
	c}) Quasar luminosity in [ergs s$^{-1}$].  The solid line and dashed
	line are reference lines showing the trend in the evolution of the
	quasar break luminosity $L^*_{\rm bol}$ (with arbitrary normalization)
	from \citet{hopkins07b}.  Note that it is the ratio of $\Delta L =
	{\rm log} (L_V / L^*_{\rm bol})$ that affects the degree of bias, so
	high-z quasars are not relatively biased despite the \lqso\ being
	somewhat higher.  Note that high-$z$ quasars are not atypically large
	in all these observables compared to low-$z$ quasars from
	\citet{kim08b}.}

    \label{fig:compare1}

\end{figure}

\subsection {The Black Hole Mass Scale in Quasars Is Tied to Normal Galaxies
Through the \msig\ Correlation}

In the context of the evolution in \mbh/\mbulge\ ratio $\Gamma$, the
discussions above on the luminosity function bias is mostly academic because
the BH mass scale in quasars is tied to normal galaxies through the \msig\
correlation \citep{onken04}.  The bias due to the LF selection is normalized
out to first order.

To second order, there are other concerns when comparing the high-$z$ sample
with low-$z$, such as the relative luminosities of the quasars, the Eddington
ratio, and the possibility that high-$z$ BHs are unusually massive.  These
concerns are addressed by Fig.~\ref{fig:compare1}, which shows that the
high-$z$ sample is not too different from the low-$z$ sample in those
respects.  The one caveat is that, even though the systematic bias in LF
selection above is normalized out to first order, there remains residual
biases relative to some reference point of the \mbb\ correlation.  Objects
more, or less, luminous compared to that reference point may lie
systematically away from the correlation, keeping in mind this is at most a
second order effect.  Taking that pivot point to be around the break of the
\phimbh, one can see in Fig.~\ref{fig:compare1}{\it c} that the high-$z$
quasar luminosities in the \citet{peng06b} sample track the evolution of the
LF break \citep[taken from][with arbitrary normalization]{hopkins07b} of the
quasars fairly closely both at low and high redshifts.

\begin{figure}[t]
    % \vspace*{-2.0 cm}
    \begin{center}
        \includegraphics[width=5.3in]{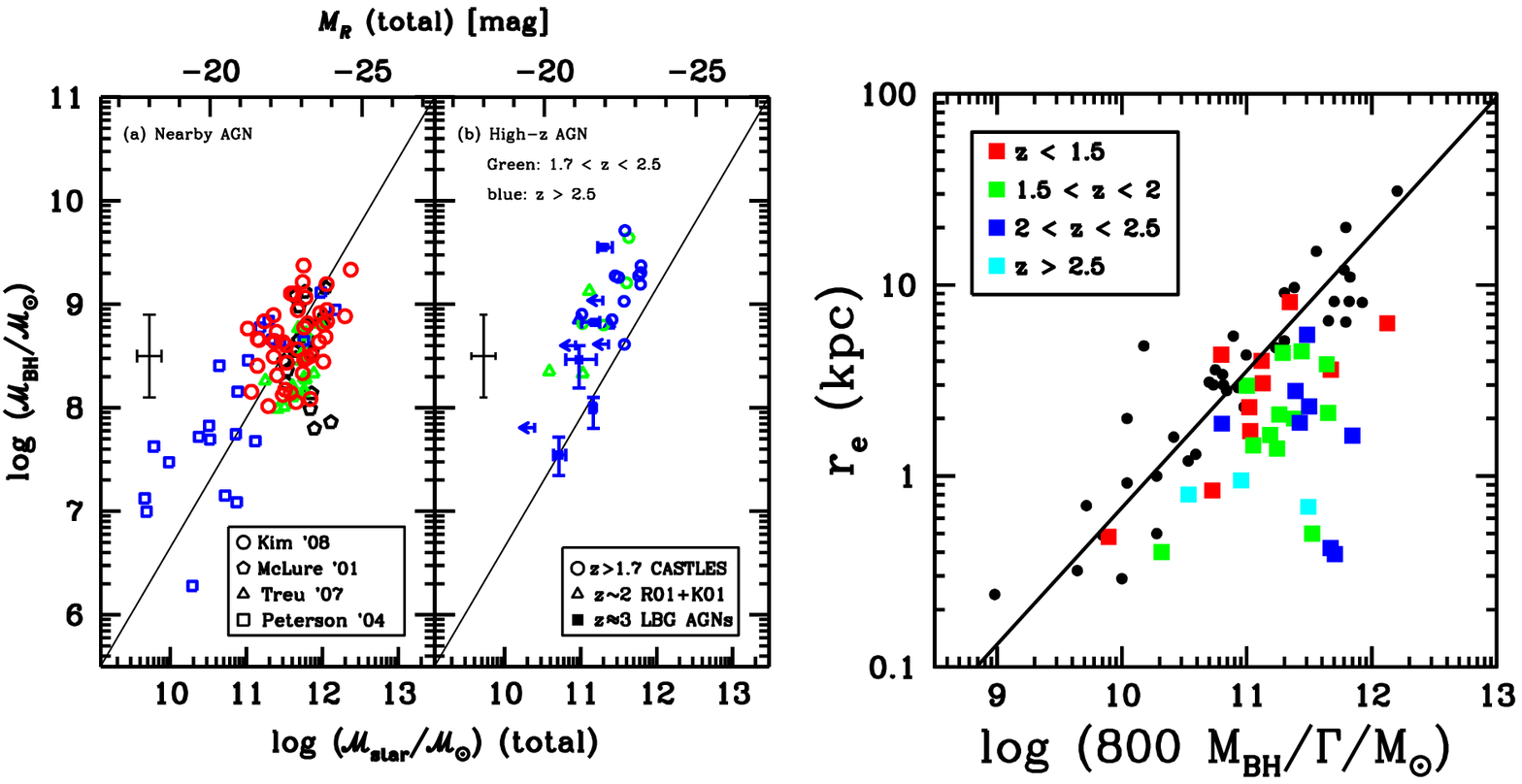} 
      % \vspace*{-1.0 cm}
    \end{center}

    \caption{(On left) {\it Comparing the \mbb\ relation for low-$z$ and
    high-$z$ quasars hosts.} {\it a}) Low-$z$ quasars.  There is no offset
    relative to normal galaxies (solid line) because the BH mass scale in
    quasars is normalized to agree.  {\it b}) However, high-$z$ quasar hosts
    exhibit an offset.}

    \label{fig:compare2}

    \caption{(On right) {\it The radius {\rm($r_e$)} vs. ``bulge mass''
	relation of high-$z$ quasar hosts compared with low-$z$ elliptical
	galaxies.} The solid circles are normal elliptical galaxies at $z=0$
	with dynamical \mbh\ measurements.  The square data points come from
	gravitationally lensed quasar host measurements of \citet{peng06b}.
	The product 800\mbh/$\Gamma$ is the expected bulge mass at the
	observed epoch as inferred from Fig.~\ref{fig:highz} or the right hand
	side of Fig.~\ref{fig:compare2}.  Quasar hosts at $z\gtrsim2$ are very
	compact.}

    \label{fig:sizes}
\end{figure}

The fact that the \mbh\ scale in quasars is normalized to normal galaxies
means that claims of \mbb\ evolution is only meaningful if low-$z$ quasars do
not show the same offset.  Figure~\ref{fig:compare2}{\it a} shows that the
low-$z$ quasars scatter around the normal galaxy correlation (solid line),
which indicates the bootstrapping does not leave large residual biases.  In
contrast, the high-$z$ sample Figure~\ref{fig:compare2}{\it b} clearly lies
off the correlation, despite the \mbh, luminosity, and Eddington ratio being
quite similar to the low-$z$ sample.

\section {Quasar Host Galaxies at $z\gtrsim2$ Are Under-Sized for Their Mass}

Another interesting evidence that high-$z$ quasar hosts have a larger
$\Gamma=$\mbh/\mbulge\ relative to $z=0$ (which can also be thought of as a
mass deficit in the bulge) comes from comparing the size vs. \mbulge\
correlation at the observed epoch with galaxies today, as shown in
Fig.~\ref{fig:sizes}.  In that Figure, the host galaxy mass is inferred from
the luminosity of the host galaxy.  But, it is useful to recast the mass in
terms of $\Gamma$, so as to emphasize how the controversial mass deficit
parameter affects the size-\mbulge\ correlation in high-$z$ quasars.  Doing
so, the host galaxy bulge mass is: \mbulge$(z) = 800 \times$ \mbh /$\Gamma$.
This equation comes from the fact that normal galaxy bulges at $z=0$ have
$\Gamma=1$ and \mbulge$(z=0)=800\times$\mbh.  The correlation of $r_e$ with
\mbulge\ is revealing because unknown luminosity selection biases are
effectively normalized away by accounting for $\Gamma$.  Figure~\ref{fig:sizes}
shows that the host galaxies at $z\approx 1$ seem to lie on the size-\mbulge\
correlation, whereas higher redshift host galaxies appear to be much more
compact per unit mass.  By $z\approx 2$, the host galaxies appear to be too
small by a factor of 2-3 compared to normal galaxies of the same mass today
\citep{peng04}.  One way to weaken the conclusions is for $\Gamma$ to be even
larger than the controversial claim, which permits these objects lie on the
modern day size-mass correlation.  The fact that massive galaxies at high $z$
appear to be very compact has been observed by a number of studies, including
\citet{trujillo06, vandokkum08, stockton08} among others, and may point to the
same evolutionary pathways between quasar host galaxies and distant red
galaxies.

\section {Conclusion}

The statistics of galaxy merging may shed some light on the controversial
finding of an evolution in the \mbh/\mgal\ ratio.  In Monte-Carlo simulations
of \citet{peng07}, high mass objects often tend to lie to the left of the
asymptotic linear correlation.  This happens because such objects were large
outliers in the initial distribution, thereby taking them longer time to
evolve onto the asymptotic relation.  Another potential explanation for the
larger \mbh/\mgal\ ratio is that the quasar phase may signify recent BH
growth, so by observing luminous quasars we catch them in a special state on
the \mbb\ correlation.  This is consistent with the simulation findings of
\citet{hopkins07a} who explain large offsets as being due to gas rich mergers
that both feed the central BH and possess a larger mass fraction in gas.  As
explained by merger statistics, the temporary up-tick in the BH mass can
subsequently merge back onto the asymptotic linear correlation through minor
mergers.  Indeed, this is seen in Monte Carlo simulations where the BHs were
artificially boosted in mass followed by regular mergers.  Combining
statistical simulations with observations that high-$z$ quasar host galaxies
are very compact, and the fact that major mergers do not change the
\mbh/\mbulge\ ratio, seem to consistently point to minor mergers being
important for transforming quasar hosts morphologically from their compact
state at $z\approx 2$ into massive, extended, elliptical galaxies today.

\bigskip

\noindent I thank Jenny Greene, Chris Kochanek, and Luis Ho for providing very
thoughtful comments.

%%%%%%%%%%%%%%%%%%%%%%%%%%%%%%%%%%%%%%%%%%%%%%%%%%%%%%%%%%%%%%%%%%%%%%%%%%%%%

\bibliography{references}

\vfill\eject

\end{document}